\documentclass[prx,twocolumn,longbibliography,english,superscriptaddress,floatfix,longbibliography]{revtex4-1}
\usepackage[thinlines]{easytable}
\usepackage{graphicx}
\usepackage{lmodern}
\usepackage{epstopdf}
\usepackage{dcolumn}
\usepackage{bm}
\usepackage{subfigure}
\usepackage{makecell}
\usepackage{amsmath} 
\usepackage[pdftex,colorlinks=true,citecolor=blue,linkcolor=blue,urlcolor=blue,bookmarks=true]{hyperref}

\usepackage{color}
\usepackage{textcomp}
\definecolor{red}{rgb}{1,0,0}

\definecolor{blue}{rgb}{0,0,1}

\definecolor{green}{rgb}{0,1,0}

\begin{document}
	\preprint{APS}
\title{Magnetic properties of a spin-orbit entangled $J_{\rm eff}$ = 1/2 honeycomb lattice }

\author{J. Khatua}
\affiliation{Department of Physics, Indian Institute of Technology Madras, Chennai 600036, India}
\author{Q. P. Ding } 
\affiliation{Ames National Laboratory, U.S. DOE, and Department of Physics and Astronomy, Iowa State University, Ames, Iowa 50011, USA}
\author{M. S. Ramachandra Rao}
\affiliation{Department of Physics, Nano Functional Materials Technology Centre and Materials Science Research Centre, Indian Institute of Technology Madras, Chennai-600036, India}
\affiliation{Quantum Centre  of Excellence for Diamond and Emergent Materials, Indian Institute of Technology Madras,
	Chennai 600036, India.}
\author{ K. Y. Choi}
\affiliation{Department of Physics, Sungkyunkwan University, Suwon 16419, Republic of Korea}
\author{A. Zorko}
\affiliation{Jo\v{z}ef Stefan Institute, Jamova c.~39, SI-1000 Ljubljana, Slovenia}
\affiliation{Faculty of Mathematics and Physics, University of Ljubljana, Jadranska u.~19, SI-1000 Ljubljana, Slovenia}
\author{ Y. Furukawa}
\affiliation{Ames National Laboratory, U.S. DOE, and Department of Physics and Astronomy, Iowa State University, Ames, Iowa 50011, USA}
\author{P.Khuntia}
\email[]{pkhuntia@iitm.ac.in}
\affiliation{Department of Physics, Indian Institute of Technology Madras, Chennai 600036, India}
\affiliation{Quantum Centre  of Excellence for Diamond and Emergent Materials, Indian Institute of Technology Madras,
	Chennai 600036, India.}

\date{\today}

\begin{abstract}
The interplay between  spin-orbit coupling, anisotropic magnetic interaction, frustration-induced quantum fluctuations, and spin correlations can lead to novel quantum states with exotic excitations in rare-earth-based  quantum magnets. Herein, we present the crystal structure, magnetization, electron spin resonance (ESR), specific heat, and nuclear magnetic resonance (NMR) experiments on the polycrystalline samples of Ba$_{9}$Yb$_2$Si$_6$O$_{24}$, in which Yb$^{3+}$ ions form a perfect honeycomb lattice without detectable anti-site disorder. The magnetization data reveal antiferromagnetically coupled spin-orbit entangled  $J_{\rm eff}$ = 1/2 degrees of freedom  of Yb$^{3+}$ ions in the Kramers doublet state. The ESR measurements reveal that the first excited Kramers doublet is 32.3(7) meV above the ground state. The specific heat results suggest the absence of any long-range magnetic order in the measured temperature range. Furthermore, the $^{29}$Si NMR results do not indicate any signature of magnetic ordering down to 1.6 K, and the spin-lattice relaxation rate reveals the presence of  a field-induced gap that is attributed to the Zeeman splitting of Kramers doublet state in this quantum material. Our experiments detect neither spin freezing nor long-range magnetic ordering down to  1.6 K. The current results suggest the presence of  short-range spin correlations in this  spin-orbit  entangled $J_{\rm eff} $ =1/2  rare-earth  magnet on a honeycomb lattice.

\end{abstract}

\maketitle
\section{Introduction}
Quantum fluctuations induced by frustration, spin correlations, and spin-orbit entanglement can stabilize exotic states in quantum materials \cite{doi:10.1146/annurev-conmatphys-020911-125138,Takagi2019,Balents2010}. Understanding emergent quantum phenomena  and associated elementary excitations is one of  the attractive pursuit  in quantum condensed matter \cite{doi:10.1146/annurev-conmatphys-020911-125138,Takagi2019,Balents2010,KHUNTIA2019165435}.  The two-dimensional geometrically frustrated triangular and  kagom\'e lattices  have been  studied extensively to realize exotic many-body quantum phenomena such as quantum spin liquid (QSL)  in condensed matter \cite{Khuntia2020,Balents2010,doi:10.1126/science.aay0668,Wen2019}.
A QSL is a highly entangled state of matter wherein spins do not exhibit long-range magnetic order even at $T\rightarrow 0$ owing to strong quantum fluctuations and intertwining of competing degrees of freedom \cite{Balents2010}. This QSL state is often characterized by exotic quasi-particle excitations such as spinons or Majorana fermions  with fractional spin quantum number \cite{ANDERSON1973153,Kitaev_2006}, which are different from the conventional
magnon excitations  with integer spin quantum numbers, as usually observed in magnetically ordered systems \cite{PhysRevLett.96.247201}.\\ The quest for such unconventional states of matter was triggered by two theoretical scenarios. The first is the resonating valance bond state proposed by P. W. Anderson
in 1973 \cite{ANDERSON1973153}.  The second one  is the Kitaev QSL on the honeycomb lattice wherein \textit{S} = 1/2  spin is predicted to fractionalize into
 emergent Majorana fermions and localized $Z_{2}$ fluxes \cite{Kitaev_2006,PhysRevLett.98.247201,doi:10.7566/JPSJ.89.012002}. 
\\
 The Kitaev model on the honeycomb lattice with $J_{\rm eff}$ = 1/2 degrees of freedom demonstrated that
the bond dependent Ising interactions provide  an alternative route to realize a frustration-driven Kitaev spin-liquid
state with Majorana fermions \cite{Kitaev_2006,Kitagawa2018}. This has sparked significant research interest in strong spin-orbit coupled 4$d$ and 5$d$  honeycomb magnets, including iridates A$_{2}$IrO$_{3}$ (A = Li$^{+}$, Na$^{+}$) \cite{PhysRevLett.108.127204,PhysRevLett.105.027204} and ruthenate $\alpha$-RuCl$_{3}$ \cite{PhysRevB.90.041112,Sears2020,Banerjee2016,Takagi2019}. 
 Beyond 4$d$/5$d$ ion based honeycomb magnets, the search for Kitaev materials has been extended to 3$d$ transition metal Co$^{2+}$ ion based honeycomb compounds such as  Na$_{2}$Co$_{2}$TeO$_{6}$ and Na$_{3}$Co$_{2}$SbO$_{6}$, in which oxygen ligands of 3$d$ ions form  nearly regular octahedra with small trigonal distortions \cite{PhysRevB.97.014407,PhysRevLett.125.047201}. Recent reports demonstrate that the combination of spin-orbit coupling and
 strong Hund’s couplings can  host
 pseudospin-1/2 degrees of freedom  with Kitaev interactions in the aforementioned cobaltates \cite{Kim_2021,PhysRevB.102.224429,Motome_2020,PhysRevB.97.014407,PhysRevLett.123.037203}.
 However, most of the 3$d$, 4$d$ and 5$d$ based honeycomb magnets show magnetic
 ordering at low temperatures due to the presence
 of inevitable defects, site disorder, and additional exchange interactions in real materials that destabilize the Kitaev QSL ground
 state  \cite{https://doi.org/10.48550/arxiv.2201.06085}.\\
  The essential ingredients to realize Kitaev spin liquid are spin–orbit entangled
 $J_{\rm eff}$ =  1/2 moments, electron correlations with bond-directional Ising exchange
 interactions, and low dimensionality \cite{TREBST20221}.  It is interesting in this respect that, many recent studies have found signatures of collective quantum phenomena in low dimensional rare-earth  magnets with anisotropic interaction between pseudospin-1/2 moments \cite{Wu2019,Li2015,PhysRevLett.117.097201,Arh2022,PhysRevB.103.064424,Hu2020,PhysRevLett.123.027201}.  \\
  Theoretically, it has been proposed that Yb-based honeycomb magnets may offer a  more faithful realization of Kitaev physics due to strong localization and spin-orbit coupling of 4$f$ electrons compared to their 4$d$ or 5$d$ counterparts \cite{PhysRevMaterials.4.104420,PhysRevB.95.085132,10.21468/SciPostPhysCore.3.1.004}. In sharp contrast to  the long-range N\'eel order state in an isotropic nearest-neighbor exchange model on the honeycomb lattice \cite{Fouet2001,PhysRevB.102.014427,PhysRevB.100.180406,Hao2020,PhysRevB.102.014427,PhysRevB.100.180406,Sala2021,https://doi.org/10.48550/arxiv.2207.02329}, the strong quantum fluctuations induced by  further nearest-neighbor frustrated exchange interaction can destabilize the N\'eel order even in bipartite honeycomb lattices \cite{PhysRevB.97.205112,Ferrari_2020,PhysRevB.96.104401,Wessler2020}.  The bipartite spin-lattice of rare-earth-based honeycomb magnets offers a promising venue to host spiral spin-liquid state with fracton quadrupole excitations \cite{Niggemann_2020,PhysRevLett.128.227201}, multiple-$q$ states in the presence of magnetic field \cite{PhysRevB.100.224404}, lattice nematic phase \cite{PhysRevB.87.024415}, and Berezinskii-Kosterlitz-Thouless phase \cite{PhysRevB.104.155139,PhysRevB.91.214412}. In this context, structurally perfect novel rare-earth 4$f$ based honeycomb magnets wherein  the combination of spin-orbit coupling and sizable crystal electric field allows for the realization of an effective spin 1/2 system with large anisotropy and strong spin correlation. These spin-orbit driven materials offer an alternate route for the experimental search for exotic quantum many-body phenomena including  Kitaev QSL \cite{10.21468/SciPostPhysCore.3.1.004,Clark2019,PhysRevMaterials.4.104420,https://doi.org/10.1002/qute.201900089,PhysRevB.106.134428,PhysRevB.104.094421,PhysRevB.103.205122,Li2021,PhysRevB.104.214410,PhysRevLett.120.207203,Arh2022,PhysRevLett.123.027201,PhysRevResearch.4.033006,PhysRevB.72.085123,PhysRevB.81.184427,PhysRevE.93.062110,PhysRevB.63.224401,PhysRevB.83.054402}. \\
Herein, we present  our results on a promising rare-earth-based quantum magnet Ba$_{9}$Yb$_{2}$Si$_{6}$O$_{24}$, where the  magnetic Yb$^{3+}$ ions form a perfect honeycomb lattice in the $ab$-plane \cite{Brgoch2013}. Magnetic susceptibility data suggest the realization of a spin-orbit entangled $J_{\rm eff}$ = 1/2 moments of Yb$^{3+}$ ions that is consistent with a low-energy Kramers doublet state at low temperature. As revealed by ESR, the ground state is well isolated from the first excited Kramers doublet, with an energy separation of 32.3(7) meV. The  Curie-Weiss fit of low-temperature magnetic susceptibility data  reveals the presence of antiferromagnetic interactions in the ground state  and detects neither spin glass nor long range magnetic ordering down to 1.9 K.
Specific heat measurements further support the absence of long range magnetic order down to 1.9 K.
 $^{29}$Si NMR measurements  in weak magnetic fields confirm the absence of long range magnetic order in the measured temperature range.  
The NMR spin-lattice relaxation rate in high-magnetic fields suggests the presence of a field-induced gap due to the Zeeman splitting of the lowest Kramers doublet state. 
Our investigation also reveals the
presence of short-range spin correlations in this rare-earth honeycomb magnet.
\section{EXPERIMENTAL DETAILS}
Polycrystalline samples of Ba$_{9}$Yb$_{2}$Si$_{6}$O$_{24}$ (henceforth; BYSO) were prepared  by a conventional solid state reaction of appropriate stoichiometry amounts of BaCO$_{3}$ (99.997 \text{\%}, Alfa Aesar), SiO$_{2}$ (99.999 \text{\%}, Alfa Aesar), and Yb$_{2}$O$_{3}$ (99.998 \text{\%}, Alfa Aesar). Prior to  use, we  preheated  BaCO$_{3}$, and Yb$_{2}$O$_{3}$ at 100$^\circ$C and 800$^\circ$C, respectively,  to remove moisture and carbonates.  All the reagents were thoroughly ground to
obtain homogeneous mixtures. The powder mixtures were  pelletized and sintered at 900$^{\circ}$C for 24 hours in air to decompose carbon dioxide. In order to obtain the desired phase, the pelletized sample was fired at 1350$^{\circ}$C for 72 hours with several intermittent grindings.  The powder x-ray diffraction (XRD) patterns were measured by employing  a smartLAB  Rigaku x-ray diffractometer  with Cu K$_{\alpha}$ radiation ($\lambda $ = 1.54 {\AA}).\\
 \begin{figure*}
	\includegraphics[width= 18.5 cm, height=5.5 cm]{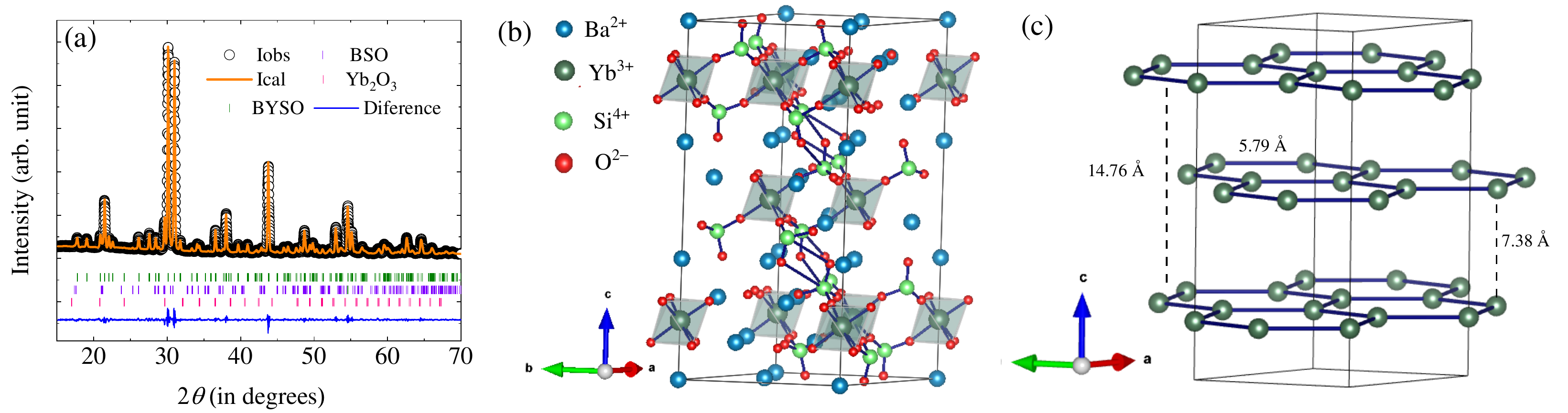}
	\caption{(a) Rietveld refinement profile of the room-temperature x-ray diffraction data of Ba$_{9}$Yb$_{2}$Si$_{6}$O$_{24}$. The black circle represents the experimentally observed data points and the orange solid line is the calculated data. The rows of vertical bars are the  Bragg reflection positions for Ba$_{9}$Yb$_{2}$Si$_{6}$O$_{24}$ (olive bars), Ba$_{2}$SiO$_{4}$ (violet bars) and Yb$_{2}$O$_3{}$ (pink bars). The blue line  is the difference
		between observed and calculated intensities. (b) A single unit cell of the trigonal crystal structure of Ba$_{9}$Yb$_{2}$Si$_{6}$O$_{24}$. The nearest-neighbor oxygen atom of Yb$^{3+}$ ions form a YbO$_{6}$ octahedra. The possible in-plane nearest-neighbor and inter-plane exchange interactions through the bridges Yb-O-Si-O-Yb, and Yb-O-Si-O-Si-O-Yb are shown, respectively. (c) Structure depicting  nearest-neighbor (5.79 {\AA}) Yb$^{3+}$ ions which form  honeycomb planes. There are three such consecutive honeycomb planes in the unit cell of Ba$_{9}$Yb$_{2}$Si$_{6}$O$_{24}$.
	}{\label{BYSO1}}
\end{figure*}
Magnetization measurements were carried out using the VSM option of Quantum Design, Physical Properties Measurement System (QD, PPMS)  in the temperature range 1.9 K $\leq \textit{T} \leq $ 300 K in magnetic fields 0 T $\leq \mu_{0}H\leq$ 7 T. Specific heat measurements were performed using QD, PPMS by thermal relaxation method, in the temperature range 1.9 K $\leq \textit{T}\leq $ 250 K in magnetic fields 0 T $\leq \mu_{0}H\leq$ 7 T. The electron spin resonance (ESR) spectrum was measured at 9.40 GHz K on a commercial X-band Bruker E500 spectrometer working at 9.40~GHz in the temperature range 4 K $\leq$ $T$ $\leq$ 140 K.
At low temperatures, the microwave power was varied between 0.01 mW to prevent signal saturation, while at high temperatures, it was set to 1 mW to avoid signal diminishment. To enhance the signal-to-noise ratio, the magnetic field was modulated with a frequency of 100 kHz and an amplitude of 0.5 mT, resulting in derivative ESR spectra. 
Field-swept $^{29}$Si ($I$ = 1/2, and gyromagnetic ratio 8.4577 MHz/T) NMR measurements  down to 1.6 K at several frequencies were carried out on a home-made phase-coherent spin-echo pulse spectrometer equipped with a 9 T Oxford magnet. NMR spectra measurements were carried out using a standard Hahn-echo  sequence while the $^{29}$Si NMR spin-lattice relaxation time was extracted from the recovery of longitudinal nuclear magnetization $M(t)$ after a time delay $t$ following a saturation pulse sequence.
\section{RESULTS}
 \begin{table}
	\vspace*{0.5 cm}
	\caption{\label{tab:table}The Rietveld refinement parameters obtained from the analysis of the XRD data taken at room temperature. The Rietveld refinements were carried out  with space group R$\bar{3}$ and yields unit cell parameters  $a$ = \textit{b} = 10.002 {\AA}, \textit{ c} = 22.127 {\AA}
		and $\alpha $ = 90$^{\circ}$, $\beta$ = 90$^{\circ}$, $\gamma $ = 120$^{\circ}$. The goodness of Rietveld refinement
		was confirmed by the following factors: $\chi^{2}$ = 3.823; and R$_{\rm exp}$= 5.25.  }
	\vspace{5 mm}
	\begin{tabular}{c c c c c  c c} 
		\hline \hline
		
		Atom & Wyckoff position & \textit{x} & \textit{y} & \textit{z} & Occ.\\
		\hline 
		Ba$_{1}$ & 3\textit{a} & 0 & 0 & 0 & 1 \\
		Ba$_{2}$ & 18\textit{f} & 0.333 & 0.666 & 0.004 & 1 \\
		Ba$_{3}$ & 18\textit{f} & 0.029& 0.668 & 0.109 & 1 \\
		
		Yb & 6\textit{c} &0 &0 &0.164 &1 \\
		Si & 18\textit{f} &0.336 &0.012 &0.073 &1 \\
		O$_{1}$ & 18\textit{f} &0.347 &0.065 &0.006 &1 \\
		O$_{2}$ & 18\textit{f} &0.480 &0.158&0.102 &1 \\
		O$_{3}$ & 18\textit{f} &-0.002 &0.173&0.107 &1 \\
		O$_{4}$ & 18\textit{f} &0.137 &0.468&0.094 &1 \\
		\hline
		
	\end{tabular}
\end{table}  
\subsection{XRD and structural details}
In order to  confirm the phase purity and obtain structural atomic parameters, Rietveld refinement of the X-ray diffraction data was performed  using the FULLPROF suite \cite{RODRIGUEZCARVAJAL199355}. The XRD results reveal that the polycrystalline samples of BYSO contain a few percentage of non-magnetic Ba$_{2}$SiO$_{4}$ and magnetic Yb$_{2}$O$_{3}$ secondary phases, which have not much effect on the overall magnetic properties of the material studied here.  In the literature, it is observed that  Ba$_{2}$SiO$_{4}$ (BSO) and Yb$_{2}$O$_{3}$ impurities are unavoidable in the polycrystalline samples of barium and silicon-based magnetic materials  \cite{Brgoch2013,PhysRevB.104.214434} as well as for some Yb-based magnets, respectively \cite{PhysRevB.107.224416,Tokiwa2021,PhysRevB.106.075132}. To quantify the percentage of the dominant and secondary phases, we performed a three-phase Rietveld  refinement.  The initial atomic coordinates were taken from refs.~\cite{Brgoch2013} and \cite{Banjare_2020} for the dominant BYSO phase and the secondary BSO phase, respectively.\\ Figure~\ref{BYSO1} (a) depicts the Rietveld refinement of the XRD data, suggesting that our polycrystalline samples contain  $\approx$ 94 \text{\%} of the dominant BYSO phase and $\approx$ 4 \text{\%} non-magnetic  BSO  and $\approx$ 2 \text{\%}  magnetic Yb$_{2}$O$_{3}$ secondary phases. The Rietveld refinement reveals that BYSO crystallizes in a trigonal crystal structure with space group R$\bar{3}$. Table~\ref{tab:table} lists the atomic parameters obtained from the Rietveld refinement.  The Yb atoms occupy only one Wyckoff atomic site 6$c$ and form a six-coordinated YbO$_{6}$ octahedron with neighboring O atoms.  The possible in-plane exchange interactions via Yb-O-Si-O-Yb super-exchange pathways is shown in Figure~\ref{BYSO1} (b). More interestingly, first nearest Yb$^{3+}$ neighbors  (5.79 {\AA}) constitutes two-dimensional honeycomb layers perpendicular to the crystallographic $c$-axis and there are three such well-separated honeycomb layers in one unit cell of BYSO (Figure~\ref{BYSO1} (c)).\\  From the structural point of view, BYSO is a bit different from the honeycomb lattice YbCl$_{3}$, which crystallizes in the monoclinic  crystal structure (space group $C2/m$ with lattice parameters $a$ = 6.732 {\AA}, $b$ = 11.620 {\AA} and $c$ = 6.328 {\AA}, $\alpha$ = $\gamma$ = 90.00$^{\circ}$  and $\beta$ = 110.551$^{\circ}$) \cite{PhysRevB.100.180406}.  However, BYSO is structurally similar to the honeycomb lattice of YbBr$_{3}$, which crystallizes in the trigonal crystal structure (space group R$\bar{3}$)   with the lattice parameters $ a = b$  = 6.971 {\AA}, $c$ = 19.103 {\AA},  $\alpha$ =  $\beta$ =  90$^{\circ}$ and  $\gamma$ =  120$^{\circ}$ \cite{Wessler2020}. Although both systems belong to the same crystal class, the ground state properties of BYSO are expected to be different due to significant differences in the ratio of $c/a$ and the exchange paths compared to YbBr$_{3}$ \cite{Wessler2020}. 
  \begin{figure*}
	\includegraphics[width= 18 cm, height= 6 cm]{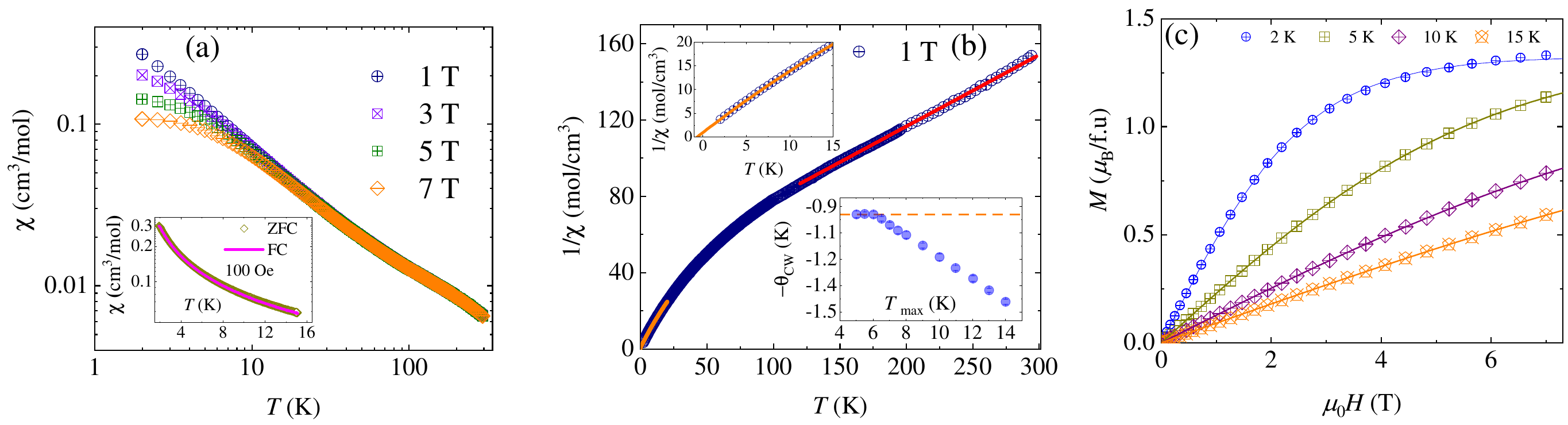}
	\caption{ (a) Temperature dependence  of magnetic susceptibility in several magnetic fields.  The inset (left bottom corner) shows the comparison of the zero-field cooled (ZFC) and
		field cooled (FC) magnetic susceptibility as a function of temperature  in a magnetic field $\mu_{0}H$ = 0.01 T.
		(b) Temperature dependence  of inverse magnetic susceptibility in a magnetic field $\mu_{0}H$= 1 T with Curie-Weiss fit in the  high-temperature (red line) and the low temperature (orange line) regions. The top inset shows the Curie-Weiss fit  of inverse magnetic susceptibility  in the low temperature region.  The bottom inset shows the estimated Curie-Weiss temperature as a function of upper limit of temperature range in the Curie-Weiss fit  where the constant value of Curie-Weiss
		temperature at low temperatures is shown by dotted orange line.
		(c) Isotherm magnetization as a function of magnetic field at several temperatures where the solid line  represents the Brillouin function fit for paramagnetic Yb$^{3+}$ spins with $J_{\rm eff}$ = 1/2 moment.
	}{\label{BYSO2}}
\end{figure*}
\subsection{\label{sec: Magnetic susceptibility }Magnetic susceptibility}
 The temperature dependence of the magnetic susceptibility of BYSO  in several  magnetic fields is shown in Figure~\ref{BYSO2} (a).
The absence of any anomaly indicates that the Yb$^{3+}$ moment does not undergo any long-range magnetic order in the measured temperature range of investigation.  The bottom inset of Figure~\ref{BYSO2} (a) depicts the zero-field cooled (ZFC) and field-cooled (FC) susceptibility  and reveals no bifurcation, which suggests the absence of spin-freezing at least above 1.9 K. \\
    Above 110 K, the inverse magnetic susceptibility data (see Figure~\ref{BYSO2} (b)) were fitted with the Curie-Weiss law, $\chi=C/(T-\theta_{\rm CW})$, where \textit{C} is the Curie constant and $\theta_{ \rm CW}$ is the Curie-Weiss temperature \cite{Bain2008}. The high-temperature Curie-Weiss fit (red line in Figure~\ref{BYSO2} (b)) yields  $\theta_{\rm  CW}$ = $-$ 111 K and effective moment $\mu_{\rm eff}$  = 4.53 $\mu_{B}$. The estimated large negative $\theta_{\rm CW}$ is attributed to the energy scale of crystal field excitations. The obtained  effective moment $\mu_{\rm eff} = 4.53 $ $\mu_{\rm B}$ is close to that expected for free Yb$^{3+}$ ions (4$f^{13}$; $L$ = 3, $S $= 1/2). 
     As Yb$^{3+}$ is a Kramers ion, generally the strong crystal electric field splits the eight-fold degenerate $J$ = 7/2 multiplet into four Kramers doublets. At low temperatures, Yb$^{3+}$ ions acquire  spin-orbit entangled $J_{\rm eff}$ = 1/2 moment.  In such a scenario, the low-temperature magnetic properties are mainly governed by the exchange interactions between the $J_{\rm eff}$ = 1/2 moment of  Yb$^{3+}$ ions in the lowest Kramers doublet state while the higher doublet states are important to understand the magnetic properties at high temperatures and high magnetic fields \cite{PhysRevResearch.3.043202}.\\ In order
     to obtain information concerning the Kramers doublet
     ground state and the nature of magnetic interactions, the Curie-Weiss fit to the low-temperature inverse susceptibility data is required. However, accurately estimating the actual Curie-Weiss temperature for rare-earth magnets is very crucial due to the high sensitivity of the fit parameters to the fitting range \cite{Arh2022}. To obtain a rough idea
     of the Kramers doublet state and dominant
     magnetic interactions between Yb$^{3+}$ moments in BYSO, the low-temperature magnetic susceptibility data were fitted with different upper temperature limit while the lower temperature limit was fixed to 4 K as described in ref. \cite{Arh2022} (see the bottom inset of Figure~\ref{BYSO2} (b)).
    Following this method, the temperature-independent fit
    parameters were found to be $\theta_{ \rm CW}$ = $-$ 0.94 $\pm$ 0.01 K and $\mu_{\rm eff} = 2.46$ $\mu_{\rm B}$. The obtained  value of effective moment, $\mu_{\rm eff} = 2.46$ $\mu_{\rm B}$ is considerably smaller  than $\mu_{\rm eff}$  = 4.53 $\mu_{B}$ expected for free Yb$^{3+}$ ions.   It implies the presence of a Kramer doublet state with $J_{\rm eff}$ = 1/2 low-energy state of Yb$^{3+}$ ions \cite{PhysRevB.102.014427}. The obtained negative Curie-Weiss temperature, $\theta_{ \rm CW}$ = $-$ 0.94 K, suggests the presence of a  weak antiferromagnetic interaction between Yb$^{3+}$ spins.\\ 
     \begin{figure*}[t]
    	\includegraphics[width=\textwidth]{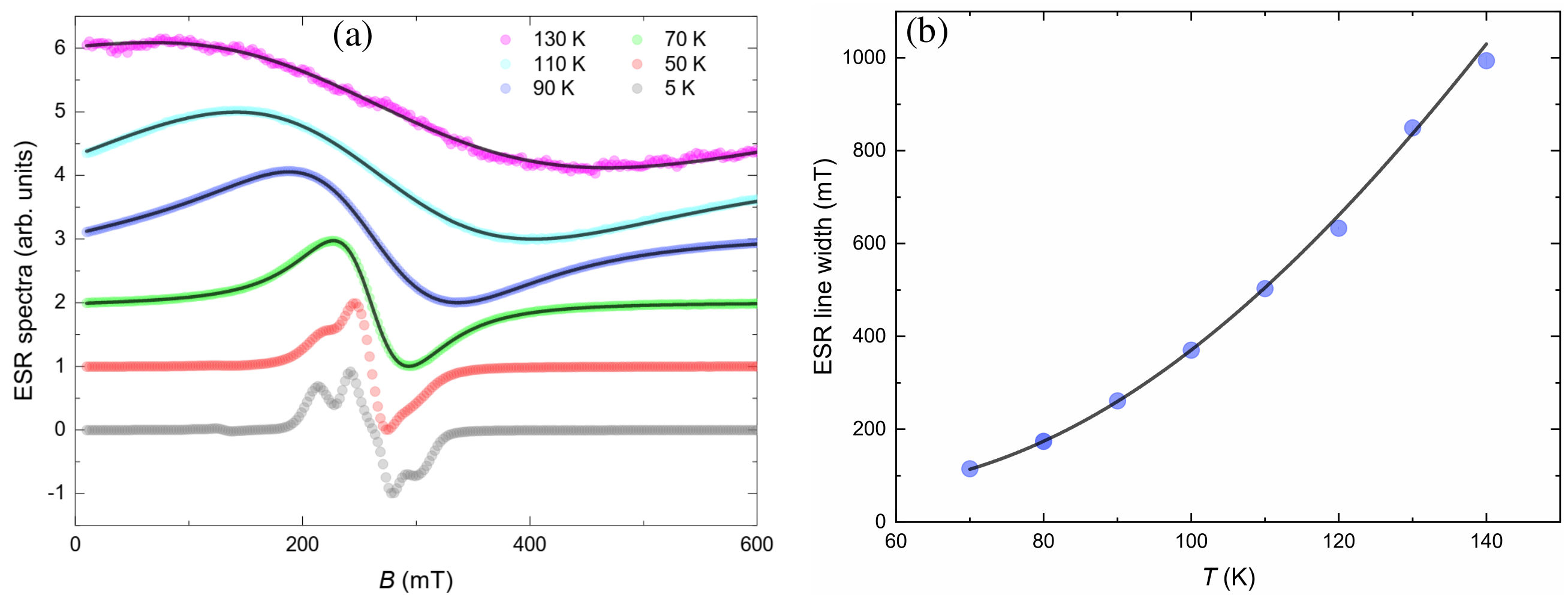}
    	\caption{(a) The ESR spectra of Ba$_9$Yb$_2$Si$_6$O$_{24}$ at various temperatures (circles), with corresponding best fits with isotropic Lorentzian line shape (solid lines) above 70~K. The spectra are shifted vertically for clarity. (b) The temperature dependence of the ESR line width (circles), with the solid line representing the fit to the Orbach model, yielding the energy gap of $\Delta = 32.3(7)$~meV (see text for details).}
    	\label{ESR}
    \end{figure*}
 Figure~\ref{BYSO2} (c) depicts the  isothermal magnetization as
a function of the magnetic field  up to 7 T  at several temperatures. For temperatures ($T\geq$ 2 K)  well above the interaction energy scale, one can model the observed isothermal magnetization following the Brillouin function, $M$/$M_{s}$ = $B_{1/2}$ ($y$),  where $B_{J}(y) = [\frac{2J+1}{2J} coth[\frac{y(2J+1)}{2J}]-\frac{1}{2J}coth\frac{y}{2J}]$ is the Brillouin function, $M_{s}$ (= g$J\mu_{B}$) is the saturation magnetization and $y=g\mu_{B}J \mu_{0}H/k_{B}T$ is the ratio of the Zeeman energy of magnetic moment to the thermal energy, $\mu_{B}$ is the Bohr magneton, and $g$ is the Lande's g-factor. The solid lines in Figure~\ref{BYSO2} (c) for 2 K, 5 K, 10 K and 15 K are the Brillouin function fit  which yields  an average Lande's g-factor, $g$ = 2.45, while $J$ was fixed to  1/2, consistent with the lowest Kramers doublet state of Yb$^{3+}$ ions in this  temperature regime.
\subsection{Electron spin resonance} 
Electron spin resonance (ESR) measurements were performed on the powder sample of BYSO at $T\geq$ 4 K. 	The ESR spectra exhibit pronounced broadening with increasing temperature, making the signal too broad for reliable analysis above 140~K (Figure\,\ref{ESR} (a)). Below 50~K, the spectra become structured, however, their line shape does not correspond to a single powder pattern due to anisotropic $g$ factors.
	Above 70~K, the spectra can be nicely fitted with a single Lorentzian curve, revealing motional narrowing effects on the line shape \cite{Zorko18}. \\The temperature dependence of the ESR line width (Figure~\ref{ESR} (b)) is due to crystal-electric-field (CEF) fluctuations, as often encountered in rare-earth magnets \cite{Arh2022}, where excited CEF levels are relatively low in energy.
	The broadening is due to the Orbach process, involving two-phonon scattering via excited CEF levels \cite{abragam2012electron}. 
	Indeed, the experimental line width $\Delta B$ is well described with the expression $\Delta B(T)$ = $\Delta B_{0}$ $+$ [$f$/(exp($\Delta/k_{B}T$) $-$1)],
	where the constant term $\Delta B_0$ arises from the magnetic anisotropy in the CEF ground state, while the exponential term describes the Orbach relaxation with $f$ being the scaling factor between the ESR line width and the spin fluctuation frequency, and $\Delta$ being the energy gap between the CEF ground state and the lowest excited state.
	The fit of the model to the experimental data yields  $\Delta B_0 = 51(3)$~mT, $f = 13(1)$~T, and $\Delta = 32.3(7)$~meV.
	This gap value is very similar to the gap of 39.4~meV observed in YbMgGaO$_4$ \cite{PhysRevLett.118.107202} or 34.8~meV observed in NaYbO$_2$ \cite{PhysRevB.100.144432}.  Both materials possess YbO$_6$ octahedra with frustrated two-dimensional arrangements, similar to BYSO.	
\subsection{Specific heat}
\begin{figure*}
	\includegraphics[width=\textwidth]{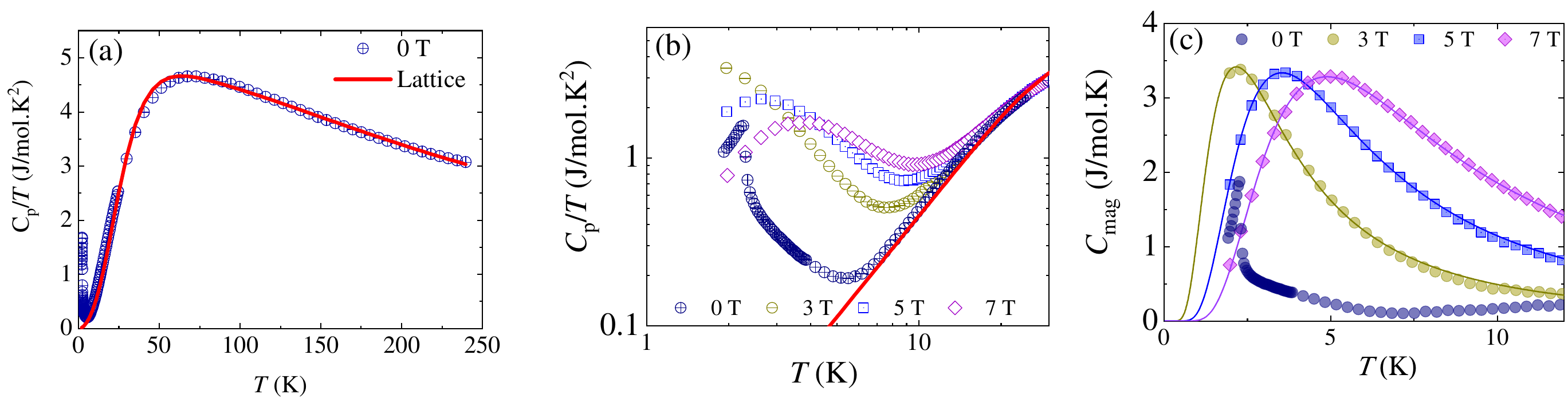}
	\caption{ Temperature dependence  of zero-field specific heat divided by temperature,  $C_{\rm p}/T$, where the red line is the fitting curve by the Debye model (see text) which accounts for phonon contributions. (b) Temperature dependence of  $C_{\rm p}/T$ in several magnetic fields. The anomaly at $\sim $ $T$ = 2.26 K in zero-field corresponds to the transition temperature of minor impurity phase of Yb$_{2}$O$_{3}$ in BYSO. (c) Temperature dependence of magnetic specific heat in several magnetic fields where the solid line depicts the fit  using Eq.~(\ref{scho}).
	}{\label{BYSO3}}
\end{figure*}

\begin{figure*}
	\centering
	\includegraphics[width=\textwidth]{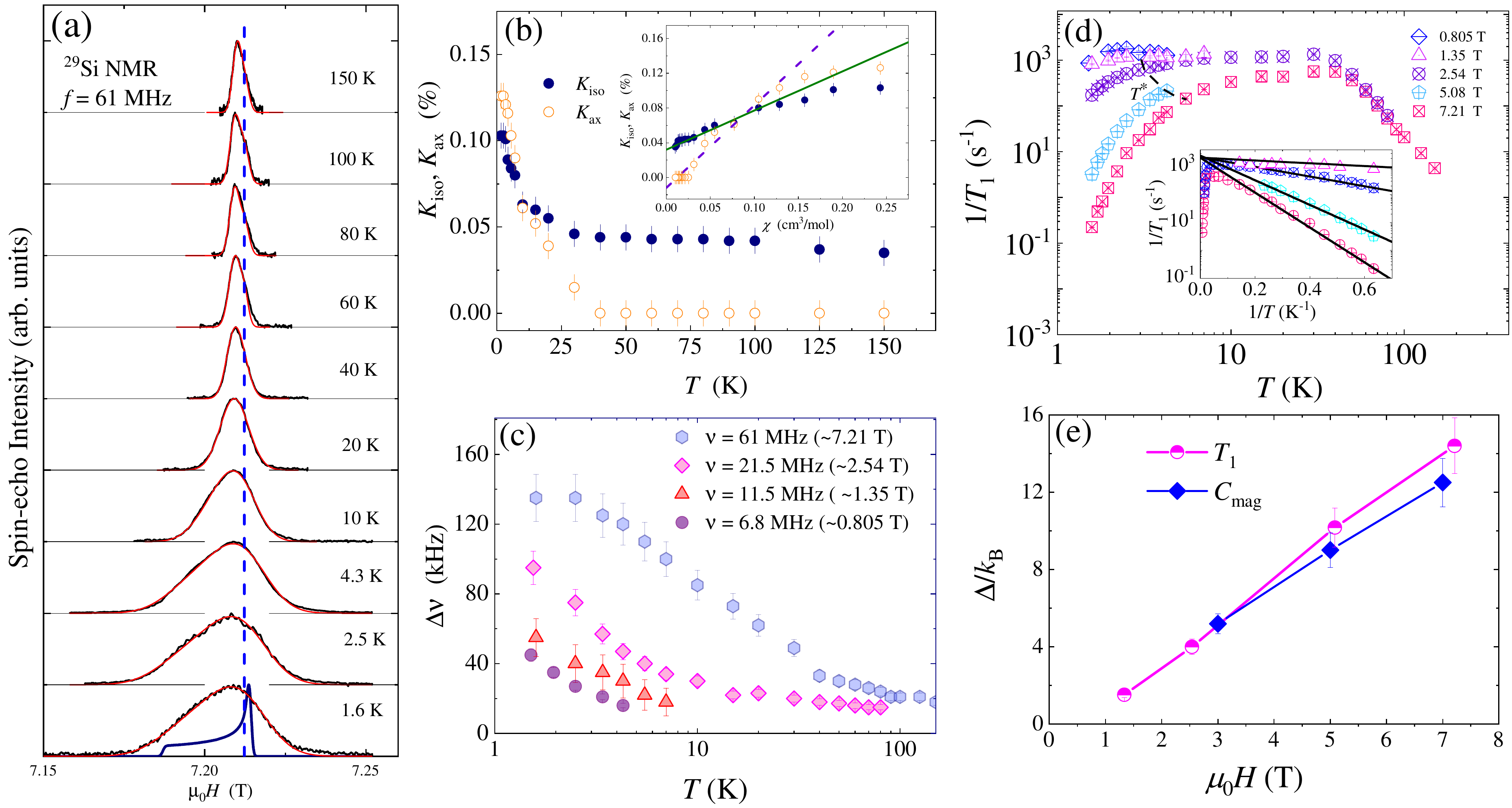}
	\caption{\label{spectra}The field-swept $^{29}$Si-NMR spectra measured at a constant frequency $\nu$ = 61 MHz at various temperatures. The blue dashed vertical line corresponds to the zero-shift reference position of 7.2123 T.  The blue curve at the bottom shows the calculated asymmetric NMR spectrum due to hyperfine anisotropy ($K_{\rm iso}$ and $K_{\rm ax}$) with nearly zero inhomogeneous magnetic broadening at $T$ = 1.6 K. The red curves are calculated spectra by taking into consideration the inhomogeneous magnetic broadening introduced by convoluting the Lorentzian function with the full width at half maximum defined as a parameter $\Delta \nu$. The slightly asymmetric shape in the spectra at higher temperatures above $\sim$ 60 K is probably due to the $^{29}$Si NMR signal at zero-shift position from the nonmagnetic impurity of Ba$_{2}$SiO$_{4}$ \cite{sm}. (b)  Temperature dependence of $K_{\rm iso}$ and $K_{\rm ax}$. The inset shows $K_{\rm iso}$ and $K_{\rm ax}$ versus magnetic susceptibility $\chi$. The solid and dashed lines are linear fits, $K_{\rm iso}$ = 0.45 $\chi$ + 0.032 and $K_{\rm ax}$ = 0.95 $\chi$ $-$ 0.012, respectively with units of \text{\%} for NMR shifts and cm$^3{}$/mol  for magnetic susceptibility.  (c) The temperature dependence of $\Delta \nu$ (inhomogeneous magnetic broadening) determined by the simulation of the NMR spectra measured at different frequencies (i.e., different magnetic field). (d) The temperature dependence of the $^{29}$Si NMR spin-lattice relaxation rate (1/$T_{1}$) measured at five different fields on a log-log scale. The inset shows the 1/$T_{1}$ as a function of inverse temperature (1/$T$) for different magnetic fields on a semilogarithmic scale. The black lines in the inset represent the fits with a phenomenological model valid for thermally activated behavior of 1/$T_{1}$ as discussed in the text. (e) Magnetic field dependence of the magnitude of the field-induced gap estimated from the 1/$T_{1}$ and heat capacity measurements.}
\end{figure*}

In order to gain further insights into ground state properties, we performed the temperature dependence of specific heat ($C_{\rm p}(T)$) of BYSO down to 1.9 K in several magnetic fields up to 7 T. The temperature dependence of zero-field $C_{\rm p}/T$ data in the temperature range 1.9 K $\leq$ \textit{T} $\leq$ 250 K is shown in Figure~\ref{BYSO3}  (a). It is noted that in zero magnetic field an anomaly appears  at $T_{\rm N}$ = 2.26 K that is attributed to the transition temperature of the unreacted  minor impurity phase of Yb$_{2}$O$_{3}$ in BYSO (Figure~\ref{BYSO3} (b)) \cite{PhysRev.176.722,10.1063/1.1709633,PhysRevB.107.224416,Tokiwa2021}. However,  the Yb$^{3+}$ moments decorating the  two dimensional honeycomb lattice do not undergo long-range magnetic order down to 1.9 K. 
\\  In order to extract the magnetic specific heat associated with Yb$^{3+}$ spins from the total specific heat,  we model the lattice contribution due to phonons using the Debye function i.e.,  $\textit{C}_{\rm lat.}(\textit{T})= 9k_{\rm B}[\sum_{n=1}^{2}C_{n}(\frac{\textit{T}}{\theta_{D_{n}}})^{3}\int_{0}^{\theta_{D_{n}}/\textit{T}}\frac{x^{4}e^{x}}{(e^{x}-1)^{2}}dx]$,
where $\theta_{D}$ is the Debye temperature, while \textit{R} and \textit{k}$_{B}$ are the molar gas constant and Boltzmann constant, respectively. 
The solid red line in Figure~\ref{BYSO3} (a) is the fitted lattice contributions that was obtained with $\theta_{D1}= $ 234 K and $\theta_{D2}$ = 456 K. In the fits, two coefficients  were fixed in the ratio C$_{D1}$:C$_{D2}$ = 17 : 24 that corresponds to the ratio of the number of heavy and light atoms in BYSO \cite{SRMN,PhysRevB.106.104408,PhysRevX.9.031005}. 
The associated magnetic contribution $C_{\rm mag}(T)$ was extracted after subtracting lattice contribution and is shown in Figure~\ref{BYSO3} (c) as a function of temperature. \\
 The temperature dependence of $C_{\rm p}(T)/T$ in magnetic fields up to 7 T is shown in Figure~\ref{BYSO3} (b).
In a magnetic field of 3 T, only a broad maximum is observed  around 2.3 K in the specific heat while the anomaly due to Yb$_{2}$O$_{3}$ is fully suppressed. This observation suggests that the low-field specific heat data most likely show a broad maximum at substantially lower temperatures due to the weak exchange coupling between $J_{\rm eff}$ = 1/2 moments of Yb$^{3+}$ ions \cite{PhysRevB.106.104404,ulaga2023quantum}. Above 3 T, this broad maximum progressively shifts toward higher
temperatures and it behaves like a Schottky anomaly, suggesting the presence of a field polarized state with
a field-induced gap in high magnetic fields \cite{10.21468/SciPostPhysCore.3.1.004}. A similar scenario has been suggested in the honeycomb lattice YbCl$_{3}$ \cite{Hao2020} and other low-dimensional frustrated magnets \cite{Wu2019,PhysRevB.106.054415}. In BYSO, the field-induced gap is attributed to the Zeeman splitting of the lowest Kramers doublet state,  effectively surpassing the intrinsic exchange interactions between $J_{\rm eff}$ = 1/2 moments in the ground-state Kramers doublet. To estimate the gap value, the high-field magnetic specific heat  data were fitted a two-level Schottky expression of the specific heat
\begin{equation}
	C_{\rm sch.}=f R \left(\frac{\Delta_{\rm s}}{k_{B}T}\right)^2\frac{{\rm exp}(\Delta_{s}/k_{B}T)}{(1+{\rm exp}(\Delta_{\rm s}/k_{B}T))^{2}},
	\label{scho}
\end{equation}   
 where $\Delta_{\rm s}$ is the gap induced by the Zeeman splitting of the ground state Kramers doublet of Yb$^{3+}$ ion, and \textit{f} measures the fraction of Yb$^{3+ }$ spins which contribute to  the splitting of the ground-state doublet. The fitted solid lines in Figure~\ref{BYSO3} (c) were obtained using Eq.\ref{scho}.  Figure~\ref{spectra} (e) displays the corresponding estimated field-induced gap. 
The estimated gap value is consistent with that obtained from  nuclear magnetic resonance spin-lattice relaxation experiments (see Figure \ref{spectra} (d)). The estimated  fraction of Yb$^{3+}$ spins was close to 1,  suggesting that almost all the Yb$^{3+}$ spins contribute to the observed effect \cite{PhysRevB.106.014409, PhysRevB.106.104408}. The spin  correlations between 4$f$ moments generally develop at low temperature in view of the weak exchange interaction between magnetic moments in these materials. This calls for sub-Kelvin thermodynamic experiments to gain deeper insights into the ground state of this 4$f$-based honeycomb antiferromagnet \cite{PhysRevB.104.L220403}. 
\subsection{Nuclear magnetic resonance}
In order to track the intrinsic static magnetic susceptibility and spin dynamics of BYSO, we also performed nuclear magnetic resonance (NMR) measurements on $^{29}$Si (nuclear spin $I$ = 1/2, gyromagnetic ratio $\gamma$$_{\rm N}$/2$\pi$ = 8.4577 MHz/T). Figure~\ref{spectra} (a) shows the field-swept $^{29}$Si NMR spectra at several temperatures at $\nu$ = 61 MHz.  The smooth evolution of the field-swept NMR spectra in the entire temperature range without developing rectangular shape or splitting of the $^{29}$Si line suggests the absence of long-range magnetic ordering in the compound \cite{doi:10.1143/JPSJ.55.1751,PhysRevB.103.214405,PhysRevB.104.115106}\\
At high temperatures, the NMR spectra are relatively narrow, but as the temperature drops, they start widening and exhibit anisotropic behavior at low temperatures. The asymmetric shapes of the spectra were well reproduced by the calculated spectra where we introduced an anisotropy in NMR shift $K$. The red curves in Figure~\ref{spectra} (a) show the calculated powder-pattern spectra with isotropic and axial components of NMR shifts ($K_{\rm iso}$ and $K_{\rm ax}$), which reproduced the observed spectra well. Here, we calculated the spectra by taking into consideration the short-range spin correlations that cause inhomogeneous magnetic broadening, which is introduced by convoluting the Lorentzian function with the full width at half maximum defined as a parameter $\Delta \nu$. The blue curve at the bottom shows the calculated asymmetric NMR spectrum due to hyperfine anisotropy ($K_{\rm iso}$ and $K_{\rm ax}$) with nearly zero inhomogeneous magnetic broadening at $T$ = 1.6 K. The NMR shift $K$ is described by $K$ = $K_{\rm iso}$ + $K_{\rm ax}$(3cos$^2\theta -1$), where $\theta$ is the angle between the principal axis of hyperfine tensor at the Si site and the external magnetic field. At higher temperatures above $\sim$ 60 K, the relatively narrow spectra also show slightly asymmetric shapes but with tails on the higher magnetic field side,  in contrast to those observed at low temperatures. From the detailed analysis, it turned out that the slightly asymmetric shape of the spectra at higher temperatures above $\sim$ 60 K can be explained by an additional $^{29}$Si NMR signal at the zero-shift position (the red curves in Figure~\ref{spectra} (a) for $T$ $>$ 60 K) which is probably due to the signal from the nonmagnetic impurity of Ba$_{2}$SiO$_{4}$ \cite{sm}.\\
Figure~\ref{spectra} (b) shows the temperature dependence of $K_{\rm iso}$ and $K_{\rm ax}$ determined by fitting the spectra. $K_{\rm iso}$ and $K_{\rm ax}$ are nearly independent of temperature within our experimental uncertainty above 40 K, however, both show a clear increase below $\sim$ 40 K beyond our experimental uncertainty.  It is interesting to point out that 40 K is close to the crossover temperature between the $J_{\rm eff}$ = 1/2 and $J$ = 7/2 states as shown in the magnetic susceptibility measurements (see Figure~\ref{BYSO2}). Below 7 K, both $K_{\rm iso}$ and $K_{\rm ax}$ saturate to finite values, whose behavior is ascribed to the strong polarization of the Yb$^{3+}$ moments in high-magnetic fields. The obtained NMR shift can be expressed as $K(T)$ = $K_{\rm 0}$ + $K_{\rm spin}(T)$ = $K_{\rm 0}$ + ($A_{\rm hf}/N_{\rm A}\mu_{B}$)$\chi(T)$, where the first term ($K_{\rm 0}$) represents the sum of temperature-independent contribution arising from the orbital susceptibility and chemical shift, the second term is the spin part of $K$ which accounts for the temperature-dependent intrinsic spin susceptibility of Yb$^{3+}$ spins. Here $A_{ \rm hf}$ is the hyperfine coupling constant, and $N_{\rm A}$ the Avogadro number. To extract the hyperfine coupling constants, the NMR shifts versus magnetic susceptibility (at 1 T) was plotted with temperature as an implicit parameter in the inset of Figure~\ref{spectra} (b) (also known as Clogston-Jaccarino plot) \cite{PhysRev.121.1357}. From the linear fit in the temperature range of $T$ = 7 - 150 K, $A_{\rm hf}$  and $K_{\rm 0}$ are estimated to be $A_{\rm hf, iso}$  = 25 $\pm$ 3 Oe/$\mu_{B}$ and $K_{\rm 0, iso}$ = 0.032 \text{\%} for the isotropic part and $A_{\rm hf, ax}$  = 53 $\pm$ 10 Oe/$\mu_{B}$ and $K_{\rm 0, ax}$ = $-$0.012 \text{\%} for axial part, respectively.\\ 
The broadening of NMR spectra with decreasing
temperature is due to the increase of the distributions in  $K_{\rm ax}$ and $K_{\rm iso}$. The temperature and magnetic field dependence of the inhomogeneous broadenings is shown in Figure~\ref{spectra} (c), where $\Delta \nu$ is plotted as a function of temperature.  $\Delta \nu$ depends on temperature and increases with increasing magnetic field. This is consistent with the magnetization measurements since $\Delta \nu$ is related to the magnetization. \\
The NMR spin-lattice relaxation rate ($T_{1}^{-1}$) probes the low-energy spin excitations related to the dynamic spin susceptibility governed by fluctuations of electron spin at the nuclear sites through hyperfine interactions. Figure~\ref{spectra} (d) depicts the temperature dependence of $T_{1}^{-1}$ with three distinct regions of different spin-lattice relaxation rates. In the entire temperature range of investigation, the relaxation rates were estimated from fitting  the recovery of longitudinal nuclear magnetization $M$($t$) by a single exponential function, $M_{z}(t)=(M_{\rm 0}-M(t))/M_{0} = A \ \ \textnormal{exp}(-t/{\textit{T}_{1}})$, where $M_{\rm 0}$ is the equilibrium magnetization, $M_{z}(t)$ is the magnetization at time $t$ after the saturation pulse, and $A$ is a constant. It is noted that all measured relaxation curves were well-fitted with a single component of $T_{1}$, implying a homogeneous distribution of spin-lattice relaxation rates in this 4$f$ honeycomb lattice system.  Upon cooling, the $T_{1}^{-1}$ increases till the plateau around 50 K in magnetic fields $\mu_{0}H$ = 2.5 and 7.2 T. This behavior is often observed in rare-earth magnets due to the depopulation of the crystal electric field and is due to the Orbach mechanism that is responsible for the ESR line broadening in BYSO \cite{PhysRevB.106.104404}.\\ 
Below 50 K, $T_{1}^{-1}$  first remains nearly temperature independent down to a field-dependent characteristic temperature, as shown by the dotted curve in Figure~\ref{spectra} (d). Although $T_{1}^{-1}$ slightly decreases with decreasing temperature in  a high magnetic field of 7.2 T, the nearly temperature-independent behavior of $T_{1}^{-1}$ in the intermediate temperature range suggests that the relaxation rate is dominated by the paramagnetic spin fluctuations of Yb$^{3+}$ spins in the crystal field ground state. One can expect the correlated magnetism of Yb$^{3+}$ spins and a field-polarized phase at low temperatures owing to the weak exchange interaction between Yb$^{3+}$ spins typical for rare-earth-based quantum magnets \cite{PhysRevB.106.104408}. The relaxation rate $T_{1}^{-1}$ decreases rapidly below the characteristic temperature $T^{*}$, suggesting that the applied magnetic field opens a gap which normally occurs when the Zeeman energy exceeds the interaction energy between Yb$^{3+}$ ions. To estimate the value of the gap in the presence of a magnetic field, we present $T_{1}^{-1}$ as a function of inverse temperature in a semi-log plot for $\mu_{0}H$ $>$ 1.35 T  as shown in the inset of Figure~\ref{spectra} (d). The solid line represents the fit to the experimental data using the phenomenological model relevant for thermally activated behavior of magnetic moments i.e., $T_{1}^{-1}$ $\propto$ exp($-\Delta_{\rm s}/k_{B}T$), where $\Delta_{s}$ is the gap value to the Zeeman splitting of the ground state Kramers doublet in the presence of a magnetic field. We find a linear variation of the gap with the applied magnetic field (see Figure~\ref{spectra} (e)), as expected when the Zeeman energy overcomes the exchange energy. The estimated $\Delta_{s}$ from the NMR measurements is in good agreement with the gap values obtained from the specific heat measurements (Figure~\ref{BYSO3} (c)). 
\section{Discussion}  
In this work, we have investigated crystal structure and ground state properties of an unexplored rare-earth based honeycomb spin-lattice Ba$_{9}$Yb$_{2}$Si$_{6}$O$_{24}$ through the combination of thermodynamic and microscopic measurements. Ba$_{9}$Yb$_{2}$Si$_{6}$O$_{24}$, nearly free from anti-site disorder, constitutes a perfect honeycomb lattice of Yb$^{3+}$ ions perpendicular to the crystallographic $c$-axis. Magnetic susceptibility data suggest the presence of  spin-orbit entangled $J_{\rm eff}$ = 1/2 moments of Yb$^{3+}$ ions with  a  Kramers doublet ground state, which are exchange-coupled by weak antiferromagnetic interactions at low temperatures. 
The mean field formula $\theta_{ \rm CW}$ = ($-$$zJS(S+1)$)/3$k_{\rm B}$ offers an approximate estimation of the exchange interaction $J/k_{\rm B}$ between $J_{\rm eff}$ = 1/2 moments of Yb$^{3+}$ ions in the $ab$-plane taking into account the nearest-neighbor $z$ = 3 and $S$ = $J_{\rm eff}$ = 1/2 for BYSO \cite{Li2020}.
 The nearest-neighbor exchange interaction in the $ab$-plane is found to be roughly 1.3 K while the
dipolar interaction energy  E$_{\rm dip}$$ \approx$ $\mu_{0}g_{\rm avg}^{2}\mu_{B}^{2}/4\pi a^{3}$ $\approx$ 0.3 \text{\%} of the nearest-neighbor exchange interactions, where $g_{\rm avg}$ is the powder average Land\'e $g$ factor and $a$ is the nearest-neighbor Yb-Yb bond length in BYSO. This implies the presence of a significant super-exchange interaction between $J_{\rm eff}$ = 1/2 moments of Yb$^{3+}$ ions in BYSO in addition to the dipolar interaction, analogous to the honeycomb lattice YbBr$_{3}$ \cite{Wessler2020}. In contrast to the corner and side sharing regular YbBr$_{6}$  octahedra in the two-dimensional honeycomb lattice YbBr$_{3}$, YbO$_{6}$ octahedra in BYSO are isolated \cite{Wessler2020}.  Thus, in BYSO, the nearest-neighbor intra-plane superexchange interaction can only occur via the Yb-O-Si-O-Yb virtual path, which may be one of the reasons for the weak antiferromagnetic interaction strength between Yb$^{3+}$ moments in BYSO. Contrarily, in YbBr$_3$, the bromine ion directly mediates the nearest neighbor superexchange interaction through the Yb-Br-Yb virtual electron hopping processes, which results in a little higher interaction strength ($\approx$ 8 K) \cite{Wessler2020}. Furthermore, in BYSO, the nature of exchange interactions between Yb$^{3+}$ moments is expected to be Heisenberg-type as the YbO$_{6}$ octahedra are connected through intermediate Si$^{4+}$ ions which prevent one of the essential requirements i.e., Yb-O-Yb  bond angles of 90$^\circ$ for stabilizing  Kitaev interactions (see Figure~\ref{BYSO1} (b)) \cite{PhysRevLett.102.017205}. \\  
Our ESR results indicate the presence of anisotropic magnetic interaction between $J_{\rm eff}$ = 1/2 moments of Yb$^{3+}$ ions. Normally, in the conventional antiferromagnetically ordered state, the crystallographic site of the probing nucleus  in an NMR experiment becomes inequivalent and senses
different magnetic fields which leads to splitting the NMR spectra when internal fields at the nucleus are parallel or antiparallel to an external magnetic field. The absence of NMR line splitting or rectangular spectra confirms the absence of long-range magnetic order in BYSO down to 1.6 K, which substantiates that the minor Yb$_{2}$O$_3$ impurity phase does not significantly affect the underlying magnetic properties of the material under consideration. The  bulk of the material maintains a dynamic ground state at least down to 1.6 K. In high magnetic fields, the exponential decay of the spin-lattice relaxation rate at low temperature is  attributed to a field-induced gap due to the Zeeman splitting of the low-energy Kramers doublet state, which typically takes place when the Zeeman energy exceeds the exchange interaction energy. Theoretically, it has been suggested that the competing magnetic interactions between nearest-neighbor and next-neighbor contacts can nevertheless lead to the realization of a quantum spin liquid state in  honeycomb lattices even in the absence of Kitaev-type interactions  \cite{Fouet2001,PhysRevB.94.214416,PhysRevB.88.165138,PhysRevLett.110.127203}. The two-dimensional honeycomb lattice  BYSO seems promising in this direction, and its  low-temperature magnetic properties on single crystal  could hold significant potential for future investigations.
 \section{SUMMARY}
 To summarize, we have synthesized and performed magnetization, specific heat, ESR, and NMR experiments on a 4$f$ electron-based material Ba$_{9}$Yb$_{2}$Si$_{6}$O$_{24}$, which crystallizes in a trigonal crystal structure with the space group R$\bar{3}$. In this material, the Yb$^{3+}$ ions constitute a perfect honeycomb lattice in the $ab$-plane. Magnetization data suggest the pseudospin-1/2 degrees of freedom of Yb$^{3+}$ ions in the Kramers doublet state, and  these $J_{\rm eff}$ = 1/2 spins  interact antiferromagnetically. The lowest CEF excited Kramers doubled is far above (at 32.3(7) meV) the ground state, as estimated from ESR. \\ The absence of long-range magnetic ordering between pseudospin-1/2 moments of Yb$^{3+}$ ions in specific heat data, which is corroborated by microscopic NMR results down to 1.6 K, suggests that the two-dimensional honeycomb lattice  Ba$_{9}$Yb$_{2}$Si$_{6}$O$_{24}$ can host a spin liquid ground state. The spin-lattice relaxation rate measurements in high-magnetic fields show the presence of a field-induced gap due to Zeeman splitting of the Kramers doublet, which is consistent with specific heat results. 
 Further studies on  single crystals of Ba$_{9}$Yb$_{2}$Si$_{6}$O$_{24}$ are highly desired to shed more insights into the anisotropic magnetic interactions, and low-energy excitations. The present family of rare-earth based honeycomb spin-lattice Ba$_{9}$$R$$_{2}$Si$_{6}$O$_{24}$ ($R$ = rare-earth ions) with distinct rare-earth elements, spin-orbit-driven anisotropy, and spin correlations provide an ideal ground to realize  exotic quantum phenomena.
 \section{Acknowledgments}
 P.K. acknowledges funding by the Science and Engineering Research Board and Department of Science and Technology, India through research grants. This research was supported by the U.S. Department of Energy, Office of Basic Energy Sciences, Division of Materials Sciences and Engineering. Ames National Laboratory is operated for the U.S. Department of Energy by Iowa State University under Contract No. DEAC02-07CH11358.
 A.Z. acknowledges the financial support of the Slovenian Research Agency through the Program No.~P1-0125 and Projects No.~N1-0148 and J1-2461.
  
\bibliographystyle{apsrev4-1}

\bibliography{BYSO_BIB}

\end{document}